\documentclass[12pt]{article}

\usepackage{amsfonts,amsmath,amsxtra}

\newcommand{\beq}[1]{\begin{equation}\label{#1}}
\newcommand\eeq {\end{equation}}
\newcommand\bqa {\begin{eqnarray}}
\newcommand\eqa {\end{eqnarray}}
\newcommand\pr {\partial}

\newcommand{\eq}[1]{eq.(\ref{#1})\ }
\newcommand{\bear}{\begin{array}}
\newcommand{\enar}{\end{array}}

\newcommand{\F}{\mathbb{F}}

\begin{document}

\def\t{\theta}
\def\T{\Theta}
\def\w{\omega}
\def\ov{\overline}
\def\a{\alpha}
\def\b{\beta}
\def\g{\gamma}
\def\s{\sigma}
\def\l{\lambda}
\def\wt{\widetilde}


\hfill ITEP--TH--64/02

\vspace{10mm}

\centerline{\Large \bf On the relations between correlation functions in }
\centerline{\Large \bf SYM/pp--wave correspondence}

\vspace{5mm}

\centerline{\bf E.T. Akhmedov\footnote{E--mail: akhmedov@itep.ru}}

\centerline{117259, ul. B.Cheremushkinskaya, 25, ITEP, Moscow}

\vspace{5mm}

\begin{abstract}
In this note we establish the explicit relation between correlation functions
in ${\cal N} = 4$ SUSY Yang--Mills theory on $S^3\times R$ in a
double scaling limit and scattering amplitudes
of the String Theory on the pp-wave background. The relation is
found for two-- and three--point correlation functions in these theories.
As a by product we formulate a dictionary for the correspondence  between
these theories: in particular, we find some unknown relations between the
parameters in these theories. Furthermore, we argue that the String Theory on
the pp--wave background is related to four--dimensional SUSY Yang--Mills theory
rather than to the reduced matrix quantum mechanics. We identify string
theory excitations which are related to the Kaluza--Klein excitation of the
Yang--Mills theory on $S^3$.
\end{abstract}

\vspace{5mm}

\section{Introduction}

It is obvious that in every scientific exploration there should be an
independent "judge". In particular, in
phenomenological theoretical physics the role of such
a judge is played by the experiment. The situation is such that
there is no any widely accepted "judge" in String Theory (ST).
Though, in this case it is natural to accept as such a judge
the "mathematical experiment". By the "mathematical experiment" we mean
a relation (formula) with LHS and RHS and equality between them,
where these sides were calculated in seemingly unrelated to each other
theories. Once someone accepted the assumptions or, if you will,
definitions underlying the calculations behind the relation in question, he can
check the relation himself and find it to be wrong or correct.

 One of the "mathematical experiments" which string theoreticians are trying
to perform nowadays is to find a relation between gauge and gravity (string) theories.
It is considered as a first step towards understanding the relation between
these two kinds of theories in general.
Recently there were found some examples of such a relation. For example:

\begin{itemize}
\item The relation between old matrix models and two-dimensional gravity (see for review
\cite{Morozov:hh}). We will refer to this relation as MM/2D gravity correspondence.

\item The relation between WZNW model and Chern-Simons theory \cite{Witten:1988hc}.
We will refer to this relation as WZNW/CS correspondence.

\item The relation between M-atrix theory and light-cone eleven-dimensional
gravity in flat space \cite{Banks:1996vh} or the relation between IKKT M-atrix
model and type IIB ST \cite{Ishibashi:1996xs}. (M--theory correspondence).

\item Relation between ${\cal N} = 4$ SUSY Yang-Mills (SYM) theory at large rank
of the gauge group and large coupling constant and type IIB supergravity
on $AdS_5\times S_5$ space with background RR tensor flux
\cite{Maldacena:1997re,Gubser:1998bc,Witten:1998qj} (see for reviews
\cite{Aharony:1999ti,Akhmedov:un,Akhmedov:1999rc}). The so called AdS/CFT correspondence.

\item Relation between ${\cal N} =4$ SYM theory in a double scaling limit and type
IIB ST on the pp-wave background \cite{Berenstein:2002jq}. The SYM/pp-wave
correspondence.
\end{itemize}

The MM/2D gravity and WZNW/CS correspondences are the relations
between small dimensional topological theories. Thus, there is no any
involved dynamics in such relations.
The M--theory correspondence already establishes a relation between
dynamical theories. SYM theory is one of the sides of the relation.
However, the SYM theory is small dimensional one, i.e. matrix quantum mechanics (MQM). Note
that the M--theory correspondence relies on SUSY non-renormalization theorems, but it
is checked for correlation functions of such operators which do not respect any SUSY.

In the context of the relation between gauge and gravity theories the AdS/CFT and SYM/pp--wave correspondences are
the most interesting among the listed above relations. They give a hope to find
an explicit ST description for a dynamical four--dimensional
gauge theory. Unlike the first three relations the AdS/CFT and
SYM/pp--wave correspondences did not receive any explicit check,
which would not rely on SUSY invariance of operators considered in the "experiment".
Let us clarify this point. The AdS/CFT correspondence is the relation of the
following kind:

\bqa \label{main}
\left\langle \exp\left\{-\, \sum_{\{j\}} \, \int d^4 x \, g^{\{j\}}(x) \,
{\cal O}_{\{j\}}\left[\Phi(x)\right] \right\}\right\rangle
\propto \Psi_{Q}\left[g^{\{j\}}(x)\right] \cong e^{{\rm i} \, I_{\rm min}\left[g^{\{j\}}(x)\right]},
\eqa
where on the LHS the average is taken
with the weight $\exp\{{\rm i} \, S_0[\Phi]\}$ in the four-dimensional
SYM theory. The content of fields on the LHS
is denoted for simplicity by $\Phi$,
${\cal O}_{\{j\}}[\Phi]$ is a basis of local gauge invariant
operators (with quantum numbers $\{j\}$) in the theory.
At the same time on the RHS of (\ref{main})
$\Psi_{Q}$ is a wave function in $AdS_5\times S_5$
supergravity characterized by concrete quantum numbers denoted by $Q$ \cite{Akhmedov:2002gq}. Note that
to find this wave function one takes the "time-slice" with respect
to the Euclidian "time" ($u$ in common notations) in $AdS_5$.

So it seems that \eq{main} is an explicit formula, which can be
checked by a direct calculation. However, the problem is that both sides of it are not
sufficiently well defined. At this stage this is a kind of the strong/week coupling relation
--- duality.
This is mainly due to the fact that one does not know how
to work with the ST on such a complicated background as $AdS_5\times S_5$. As the result almost
anything, which can be calculated on the RHS is non-analytic in the coupling constant
of the LHS or independent of it. In fact, normally any
calculation which can be performed on the
RHS is a strong coupling calculation from the SYM point of view.
Thus, all observations that favor the AdS/CFT correspondence strongly
rely on correlation functions of SUSY respecting operators.
This is the main obstacle on the way to perform the "mathematical
experiment" for the AdS/CFT correspondence.

On the other hand, in the SYM/pp--wave correspondence the ST on the pp--wave background
is a rather simple theory \cite{Metsaev:2001bj,Metsaev:2002re}. This gives a hope to check the
correspondence explicitly. In fact, many quantities which can be calculated on the ST
side appear to be analytic in the SYM coupling constant \cite{Berenstein:2002jq}.
This means that the correspondence can be established for any correlation functions
in the theories without any respect of SUSY.
However, to our knowledge, for the SYM/pp--wave correspondence there is no any well established
explicit relation like \eq{main}. We believe, however, that it should exist because the correspondence
was obtained as a limit of AdS/CFT correspondence \cite{Berenstein:2002jq}.

In the SYM/pp--wave correspondence
it is established that the masses of the ST states
are equal to the conformal dimensions of the SYM operators
\cite{Berenstein:2002jq,Gross:2002su,Santambrogio:2002sb}.
Furthermore, there is a relation between the structure constants
for operator algebras in both theories \cite{Constable:2002hw,Spradlin:2002rv}. However, there was not found any
explicit relation between correlation functions in SYM and ST.
Moreover, it is not clear whether the ST on pp-wave is related to the
four--dimensional SYM or only to the reduced MQM.
So the "experiment" is not yet complete in this case. In contrast to AdS/CFT correspondence it is not
even clear what should be checked.

The goal of this paper is to formulate the SYM/pp--wave correspondence
in the form similar to \eq{main} and to make a few rigorous checks.
We establish the explicit relation between space--time dependencies
of the two-- and three--point correlation functions in
SYM theory and ST on the pp--wave background.
We present arguments which support that it is four--dimensional SYM rather than
just reduced MQM is related to the ST.
As a by product we formulate the vertex operator formalism in the light-cone (LC) gauge
for the conformal field theory describing ST on the pp-wave background.

The explicit formula which we obtain for the SYM/pp--wave
correspondence looks as
follows. On the SYM side we consider generating functional of connected correlation functions of
the ${\cal O}^{J\, n}_{L,\, k,\, m}(t)$ operators (and their cousins), where

\bqa
{\cal O}^{J\, n} \left(t,\, {\vec \phi}\right) = \sum_{L,\, k, \, m} \,
{\cal O}^{J\, n}_{L,\, k,\, m}(t) \, Y_{L,\, k,\, m} \left({\vec
\phi}\right).
\eqa
In this formula ${\vec \phi}$ is the coordinate on
$S^3$, $Y_{L,\, k,\, m} \left({\vec \phi}\right)$
are spherical functions on $S^3$ and ${\cal O}^{J\, n} \left(t,\, {\vec
\phi}\right)$ are operators which were
considered in \cite{Berenstein:2002jq}. The generating functional in question is:

\bqa
W_{SYM}(g) = \log \int \exp\left\{{\rm i}\, S_0 + {\rm i} \int \, g_0(t) \, {\cal O}(t) +
{\rm i} \int \, g^n_{L,\,k,\,m}(t)\, {\cal O}^{J \,n}_{L,\,k,\,m}(t) + \dots\right\},
\eqa
where the dots are standing for the cousins of ${\cal O}$ which survive in the double
scaling limit of \cite{Berenstein:2002jq}. The second term in the exponent on the RHS is due to the
chiral operator defined in \cite{Berenstein:2002jq}.

At the same time, on the ST side we consider generating functional of the
mass--shell correlation functions of the conformal operators $V^{p^+\, n}_{\vec M}$ (which we define
below):

\bqa
Z_{ST}(g) = \int \exp\left\{{\rm i} \, S_{st} + {\rm i} \, \int g^n_{\vec
M}(p^-) \, e^{{\rm i}\, p^- \, x^+ + {\rm i} \, p^+\, x^-} \, V^{p^+ \, n}_{\vec M} +
\dots\right\},
\eqa
where ${\vec M}$ is a vector taking values in eight--dimensional lattice.

We establish relations between the quantum numbers of ${\cal O}^{n\, J}_{L,\,k,\,m}(t)$ and
of $V^{p^+ \, n}_{\vec M}(p^-)$ operators: three components of the eight--vector ${\vec M}$
are related to the Kaluza--Klein (KK) quantum numbers $L,\,k,\,m$.
As well we show that the scale parameter of the pp--wave metric --- $\mu$
in the notations of \cite{Berenstein:2002jq} --- is related to the inverse radius of the
$S^3$ on the SYM side. One can see similarity
between these two parameters already at the first glance: in both theories
we can scale them to any fixed non-zero value.

Furthermore, we show that

\bqa
g_{\dots} (t) \propto \int dp^- g_{\dots}(p^-) e^{{\rm i} \, p^- \, t}.
\eqa
Hence, we argue that there should be a relation as follows:

\bqa
\left. \left(\frac{\delta^K}{\delta g_{\dots}^K(p^-)} W_{SYM}(g) \right) \,\,
\left/\phantom{1^{\frac12}}\right.
\left( \frac{\delta^K}{\delta g_0^K(p^-)}
W_{SYM}(g)\right) \right|_{g_0 = 0, \, g_{\dots} = 0} = \left. \frac{\delta^K}{\delta g_{\dots}^K(p^-)}
Z_{ST}(g)\right|_{g_{\dots} = 0}. \label{main1}
\eqa
The normalization (denominator) on the LHS is chosen so that to make the correlation
function well defined in the double scaling limit, i.e. to cancel the part in
the conformal dimension
which is due to the multiplicity of the fields standing inside ${\cal O}$'s.

In this note we check the relation (\ref{main1}) on the planar level for
two-- and three-- point correlation functions. But we believe that it could
be established at any level.

\section{String Theory side}

In this section we discuss ST on the pp--wave background. To keep
the paper self--contained we start with the derivation of the pp-wave metric from the "AdS" one.
Then we proceed with the LC quantization of the strings on the pp-wave
background. We conclude this section with the derivation of the string vertex operators
and calculations of two-- and three--point mass-shell amplitudes of the ST.

\subsection{The pp--wave metric as the limit of $AdS_5\times S_5$}

To establish the SYM/pp--wave correspondence one assumes that the
AdS/CFT correspondence is correct in its strong
form\footnote{One assumes that SYM theory is related to the full ST on $AdS_5\times S_5$ rather
than just to its supergravity limit.}
and takes a kind of the double scaling limit on both sides of the relation.

On the AdS side one considers type IIB ST on the $AdS_5\times S_5$
background (in this text we always use Minkowski signature):

\bqa
ds^2 = R^2 \left[- dt^2 \, {\rm ch}^2(\rho) + d\rho^2 + {\rm
sh}^2(\rho) \, d{\Omega'}_3^2\right] + R^2 \left[d\psi^2 \,
\cos^2(\theta) + d\theta^2 + \sin^2(\theta)\,
d\Omega_3^{2^{\phantom{\frac12}}}\right], \nonumber \\
- \infty < t < + \infty \label{metr1}
\eqa
with a flux of the RR
four--form gauge field. Here $R^4 \propto \alpha' g_s N$ is the
radius of both $AdS_5$ and $S_5$ expressed in terms of the string
coupling constant $g_s$ and the RR flux $N$; $\alpha'$ is the
inverse string tension. Note that the boundary of this metric is
at $\rho\to \infty$ and is $S^3 \times R$ parameterized by $t$ and
$\Omega'_3$.

To obtain the pp-wave limit one changes the coordinates to:

\bqa
x^+ = t; \quad x^- = R^2 \frac{t-\psi}{2}; \quad \rho =
|{\vec r}|/R; \quad \theta = |{\vec y}|/R,
\eqa
where ${\vec r}$
and ${\vec y}$ are four-vectors. At the end, one takes $R\to
\infty$ while keeping $x^{\pm}, r$ and $y$ finite. As the result
one obtains from \eq{metr1} the pp-wave metric \cite{Blau:2002dy}:

\bqa
ds^2 = - 4 dx^+\,dx^- - \mu^2 \, x_a^2 \left(dx^+\right)^2 + dx_a^2, \label{metr2}
\eqa
where $x_a = ({\vec r}, {\vec y})$, $a = 1, ...,8$.
In the limit in question the RR flux is:

\bqa
F_{+1234} = F_{+5678} = const \cdot \mu \label{RR}
\eqa
with an arbitrary dimensionfull parameter $\mu$.

To have a reasonable theory in this limit one must keep the light-cone energy
$p^-$ and momentum $p^+$ of the strings finite:

\bqa
H_{LC} = 2p^- & = & - p_+ = {\rm i} \, \pr_{x^+} = {\rm i} \, \left(\pr_t +
\pr_\psi\right) = \left(\Delta - J\right) \, \mu, \nonumber \\
2p^+ & = & - p_- = - \frac{\tilde{p}_-}{R^2} = \frac{\rm i}{R^2} \,
\pr_{\tilde{x}^-} = \frac{\rm i}{R^2}\, \left(\pr_t - \pr_\psi\right)
= \frac{\Delta + J}{R^2\, \mu}, \label{delta}
\eqa
where $\Delta$ is the energy in $AdS_5\times S_5$ ST,
while $J$ is the angular momentum along $\psi$ direction or charge under $U_R(1) \subset
SU_R(4) \cong SO_R(6)$ --- group of rotations of the six coordinates transversal to the D3--branes.
The D3--branes are the source for the $AdS_5\times S_5$ space.
Thus, to keep the quantities in
question finite in the $R\to\infty$ limit one must take the double scaling limit

\bqa
R \to \infty, \quad (g_s \, N\to\infty), \quad \Delta \sim J \to \infty, \quad {\rm and} \quad J^2/R^4 \propto
\frac{J^2}{g_s \, N} = const. \label{limit}
\eqa
As a side remark relevant for our further discussion
let us note that according to \cite{Akhmedov:2002gq,Akhmedov:1998vf} the AdS/CFT correspondence
tells us that the SYM theory lives
on any hyper--surface $\rho = const$ of (\ref{metr1}). The latter constant defines the energy
scale of the SYM \cite{Maldacena:1997re,Akhmedov:2002gq,Akhmedov:1998vf}.
Hence, after the double scaling limit is taken the SYM theory lives at any four dimensional
hyper--surface $r = const$ of the space (\ref{metr2}), which is parameterized by coordinates
$t$ and $\Omega'_3$. Moreover, according to \cite{Berenstein:2002sa} the time coordinate $t$ of the SYM
theory should be identified with $x^+$ --- the time coordinate of
the ST on the pp--wave background. We come back to this point below.

\subsection{Quantization of strings on the pp--wave background}

The LC ST action in the pp--wave background in the GS formalism is
\cite{Metsaev:2001bj,Metsaev:2002re}:

\bqa
S_{LC} = \frac{1}{2\pi\alpha'} \, \int d\tau \, \int_0^{2\, \pi\, \alpha' \,
p^+} d\sigma \left\{\pr x_a \pr x_a - \mu^2 \,  x_a^2 -
\bar{S}\, \left(\hat{\pr} + \mu \, \Gamma^{1234}\right)\,
S\right\} = \nonumber \\
= \frac{1}{2\pi} \, \int d\tau \, \int_0^{2\, \pi} d\sigma \left\{ \left(\pr_\tau
x_a^{\phantom{\frac12}}\right)^2 -
\frac{\left(\pr_\sigma x_a\right)^2}{(\alpha'\, p^+)^2} - \mu^2 x_a^2 -
\bar{S}\, \left({\rm i} \, \hat{\pr}_\tau -
\frac{\hat{\pr}_\sigma}{\alpha'\, p^+}  + \mu \, \Gamma^{1234}\right)\,
S\right\},\label{LCact}
\eqa
where $x_a$ are two-dimensional scalars describing embeddings of the strings
into the target-space; $S$ are two-dimensional Majorana fermions, which
carry positive chirality spinor indexes under the $Spin(8)$ group.
The $x$ and $S$ fields in the second line differ from those in the first
line by the rescaling by $\sqrt{p^+}$.

The bosonic part of this action appears from the non--linear $\sigma$--model
with the pp--wave metric, while the mass term for the fermions can be understood as appearing from the
coupling of the GS fermions to the RR background (\ref{RR}):
$\Gamma^{1234}$ is the corresponding anti-symmetric product of the
gamma-matrixes.

 The harmonic expansion of $x_a(\sigma, \, \tau)$ is \cite{Metsaev:2001bj,Metsaev:2002re}:

\bqa
x^a(\sigma,\, \tau) = \cos{\mu\tau} \, x_0^a + \frac{1}{\mu} \, \sin{\mu\tau}
\, p^a_0 + {\rm i} \, \sum_{n \neq 0} \frac{1}{\omega_n}\,\left( b_n^a \exp\left\{-
{\rm i} \, \left( {\rm sign}(n) \, \omega_n \, \tau - \frac{n\, \sigma}{\alpha'\,p^+}\right)
\right\} + \right. \nonumber \\ \left. +
\tilde{b}_n^a \exp\left\{-
{\rm i} \, \left({\rm sign}(n) \, \omega_n \, \tau + \frac{n\, \sigma}{\alpha'\,p^+}\right)
\right\}\right), \label{expansion}
\eqa
where $0 \leq \sigma \leq 2\, \pi\, \alpha' \, p^+$ and

\bqa
\omega_n = \mu \, \sqrt{1 + \frac{n^2}{(\mu \alpha' \, p^+)^2}}. \label{states}
\eqa
After quantization we get the following commutation relations:

\bqa
\left[b^a_n, \, b^c_m\right] = \frac{\rm i}{2}\omega_n \delta_{m+n, 0} \,
\delta^{ac}, \quad \left[\tilde{b}^a_n, \, \tilde{b}^c_m\right] =
\frac{\rm i}{2}\omega_n \delta_{m+n, 0} \,
\delta^{ac}, \quad \left[b^a_n, \, \tilde{b}^c_m\right] = 0. \label{commut}
\eqa
Note that in the $\mu\to 0$ limit $b$ and $\tilde{b}$ become standard left
handed (holomorphic) and right handed (anti-holomorphic) harmonics and \eq{expansion}
becomes the standard expansion.

 At the same time the Laplace equation (both for scalars and for tensors) in the pp-wave
metric is:

\bqa
\left(- 2^{\phantom{\frac12}} \pr_+\, \pr_- - x_a^2 \, \pr_-^2 +
\pr_a^2 + 2 \, {\rm i} \, c \, \pr_- - m^2 \right) \F(x) = 0,
\eqa
where $c$ is the specific constant \cite{Metsaev:2001bj,Metsaev:2002re} ($c = 0$ for scalars)
and $m$ is the mass of the field. The solution of this equation is:

\bqa
\F(x) \propto e^{{\rm i}\, p^+\, x^- + {\rm i} \, p^-\, x^+} \, \chi_{\vec M}(x,\, p^+),
\quad p^- = - c - \left(\frac{8}{2} + \sum_{a=1}^8 M_a\right) - \frac{m^2}{2\, p^+}
\eqa
where $\chi_{\vec M}(x) = \prod_a \chi_{M_a}(x)$
are the wave functions of the eight-dimensional harmonic oscillator.
In particular, $\chi_0(x, \, p^+) = \exp\{-p^+\, x_a^2/2\}$.
Note that the vector ${\vec p}, \, p^+, \, p^-$ in the flat space--time is
exchanged for ${\vec M}, \, p^+, \, p^-$ in the pp--wave background.

  Thus, in the pp--wave background
the string center of mass degrees of freedom are in a particular
oscillator quantum state \cite{Metsaev:2001bj,Metsaev:2002re}:

\bqa
|0, p^+\rangle \quad {\rm or} \quad
\prod^8_{a=1}\,\left(b^+_{0\, a}\right)^{M_a}\, |0, p^+\rangle = |{\vec M}, p^+\rangle,
\quad M_a \in {\mathbb Z}
\nonumber \\ {\rm where} \quad \langle x|{\vec M}, p^+\rangle = \chi_{\vec M}(x,
p^+),
\eqa
and $b^{\pm}_0 \propto x_0 \pm {\rm i}\, p_0/ \mu$.

Furthermore, above each of these states there is the tower of the
string excitations created by the action of the $b^+_n$ operators
with $n \in Z$, $n\neq 0$.
The mass levels of the ST are given by $\sum_n \omega_n \, N_n$ \cite{Metsaev:2001bj,Metsaev:2002re}
and the constraint on the physical states is:

\bqa
P = \sum_{-\infty}^{+\infty} n \, N_n = 0,\label{cons}
\eqa
where $P$ is the two--dimensional momentum operator and $N_n$ is the
degeneracy of the $n$-th level \cite{Metsaev:2001bj,Metsaev:2002re}.

\subsection{Vertex operators and correlation functions}

Let us consider physical states in the theory. Due to the condition
(\ref{cons})
the simplest physical string excitations are given by:

\bqa
b^{+a}_n \tilde{b}_n^{+c} |{\vec M}, p^+\rangle, \label{physst}
\eqa
where $b^+_n = b_{-n}$, $n > 0$.
The mass of such a state (above the zero--energy level of the eight--dimensional oscillator) is:

\bqa
m_{{\vec M}, \,n} = \mu \left(\sum^8_{a=1} M_a + 2 \, \sqrt{1 +
\frac{n^2}{(\mu \, \alpha' \, p^+)^2}} \right). \label{maphst}
\eqa
Let us construct LC conformal vertex operator for creation of such a state.

  It is not hard to see that:

\bqa
\int d\tau d\sigma \, \left[\left({\rm i} \,
{\rm sign}(\pr_\sigma) \, \sqrt{\mu^2 - \pr^2_\sigma} + {\rm i} \,\pr_\tau
\right)^n x^a \right]\, \left[\left({\rm i} \,
{\rm sign}(\pr_\sigma) \, \sqrt{\mu^2 - \pr^2_\sigma} - {\rm i} \,\pr_\tau
\right)^n x^c\right] \, e^{ - 2 \, {\rm i} \, \omega_n \tau} \, \, |0 \rangle \propto \nonumber \\
\propto b^{+a}_n \tilde{b}_n^{+c} |0 \rangle,
\eqa
where $|0\rangle$ is the vacuum annihilated by $b$'s without $+$ and ${\rm sign}(\pr_\sigma)$ is the sign of
the corresponding derivative acting on harmonics of $x$.
In the limit $\mu\to 0$ such a vertex operator becomes the standard vertex operator for
the strings in the flat space:

\bqa
\int d\tau d\sigma \, \pr^n_+ x^a \, \pr^n_- x^c \, e^{ - 2\, {\rm i}\, n \, \tau},
\eqa
where $\sigma$ and $\tau$ are coordinates on the cillinder\footnote{Note that if one were
using coordinates on the complex plane the exponent under the integral in this expression
would be absent.} and $\pr_{\pm} = \pr/\pr\sigma \pm \pr/\pr\tau$.

 Thus, the local vertex operator for the scalar part of the state (\ref{physst}) is given by:

\bqa
V^{p^+\, n}_{{\vec M}}[x(\sigma,\, \tau)] \propto \left[x_a
\, \left(\pr^2 + \mu^2 \right)^{n\phantom{\frac12}} x_a \right] \,
e^{- 2 \, {\rm i} \, \omega_n \tau} \, e^{{\rm i} \, p^+ \, x^- +
{\rm i} \, p^- \, x^+} \, \chi_{\vec M}(x, \,
p^+),\label{vertexloc}
\eqa
where flat space factor $e^{{\rm i} \,
{\vec p} \, {\vec x}}$ is exchanged for $\chi_{\vec M} (x, \,
p^+)$. The mass-shell condition is:

\bqa
p^- = m_{{\vec M}, \,n},
\eqa
which is imposed to respect conformal invariance.

  Let us calculate mass-shell correlation functions for such operators.
The K-point function is given by:

\bqa
\left\langle \int d\sigma_1 d\tau_1
\overline{V}^{p^+_1}_{{\vec M}_1, \, n_1}(\sigma_1,\, \tau_1) \,
\int d\sigma_2 d\tau_2 \overline{V}^{p^+_2}_{{\vec M}_2, \,
n_2}(\sigma_2,\, \tau_2) \, \dots \right. \nonumber \\ \left.
\dots \, \int d\sigma_{K-1} d\tau_{K-1} V^{p^+_{K-1}}_{{\vec
M}_{K-1}, \, n_{K-1}} (\sigma_{K-1},\, \tau_{K-1}) \int d\sigma_K
d\tau_K V^{p^+_1}_{{\vec M}_K, \, n_K}(\sigma_K,\, \tau_K)
\right\rangle_{MS} = \nonumber \\ = \delta\left(\sum^K_{i=1}
p^+_i\right) \, \prod^K_{j=1} \delta\left(p^-_j -  m_{{\vec M},
\,n}\right) \left\langle {\rm CFT}\right\rangle_K \label{stcorr1}
\eqa
where the $\delta$--functions for $p^-_j$ appear from the
mass-shell condition; $\left\langle {\rm CFT}\right\rangle_K$ is
the standard K--point LC correlation function in ST, which is
independent of all $p^-_j$. It is a rather complicated correlation
function due to the presence of the factors $\chi_{\vec M}(x, \,
p^+)$ \cite{Tseytlin}. Such a correlation function is necessary to
study to understand the relation of this ST to SYM in full detail
and, in particular, to understand the structure of the algebra of
the vertex operators (\ref{vertexloc}).

Two--point CFT correlation function is easy to calculate\footnote{Because the $\chi$ factors in
\eq{vertexloc} can be absorbed into the brackets: $\langle \chi_0(x,\, p^+) \dots \chi_0(x, \, p^+)\rangle
= \langle 0, \, p^+| \dots |0, \, p^+\rangle = \langle \dots \rangle$. At the same time
\bqa
\left\langle \chi_{\vec M}(x,\, p^+) ...
\chi_{\vec M}(x, \, p^+)\right\rangle = \left\langle 0, \, p^+|\prod_a \left(b^a_0\right)^{M_a} ...
\prod_c \left(b^{c+}_0\right)^{M_c}|0, \, p^+\right\rangle = \nonumber \\ =
\left\langle \int \prod_a\left(
e^{{\rm i} \, \mu\, \tau} \, x_a\right)^{M_a} \dots \int \prod_c\left(
e^{- {\rm i} \, \mu\, \tau} \, x_c\right)^{M_c}\right\rangle. \nonumber
\eqa}. It is proportional to

\bqa
\langle CFT \rangle_2 \propto \prod_{a = 1}^8 \delta\left(M_a^{(1)}
- M_a^{(2)}\right) \cdot \delta_{n_1\, n_2} \dots \propto \delta_{p^-_1, \, p^-_2} \dots
\,. \label{stcorr2}
\eqa
It is not hard to see that the two--point function is nothing but the
following transition amplitude in the ST

\bqa\label{res} \frac{\left\langle {\vec M}_1, p^+_1 \right| \,
b_{n_1} \, \tilde{b}_{n_1} \, e^{{\rm i} \, H_{LC} \, \Delta x^+}
\, b^+_{n_2} \, \tilde{b}^+_{n_2} \left|{\vec M}_2,
p^+_2\right\rangle}{\left\langle {\vec M}_1, p^+_1 | {\vec M}_1,
p^+_1 \right\rangle} \eqa where $H_{LC}$ is the Hamiltonian
corresponding to the action (\ref{LCact}) expressed in terms of
the harmonics $b$ of the field $x(\sigma, \tau)$.

At the same time, the three--point correlation function is
proportional to

\bqa
\langle CFT \rangle_3 \propto \delta_{p^-_1, \, p^-_2 + p^-_3} \dots \,.
\label{stcorr3}
\eqa
Calculation of what is standing instead of dots we leave for future. The three--point correlation functions give OPE
coefficients which were considered in \cite{Constable:2002hw,Spradlin:2002rv} in a different approach.
For our purposes, however, we need only space--time dependence of the string
scattering amplitudes.

  Thus, to find a correspondence of the ST in question to the SYM theory
we should reproduce such correlation functions from the SYM. This is exactly what we are
doing below.

\section{SUSY Yang-Mills theory side}

We start with the identification of the relations between different
parameters in the SYM and ST. From \eq{delta} it follows that:

\bqa
\alpha' \, p^+ \, \mu \propto \frac{J}{\sqrt{g_s \, N}}. \label{relation1}
\eqa
From the ABC of the AdS/CFT correspondence we know that $\Delta$ is
the energy of a state in the SYM theory on $S^3\times R$, $J$ is $U_R(1)$ charge,
where $U_R(1) \subset SU_R(4) \cong SO_R(6)$. The $SO_R(6)$ is the group of
rotations of the six scalars present in the ${\cal N} = 4$ SYM theory, while
$SU_R(4)$ rotates four Weil fermions. There is the unitarity bound in the
SYM theory which states that: $\Delta \geq J$. Furthermore, $N$ is the rank of the
gauge group. As well it is known from the $D3$--brane action that:
$g_s = const \, g^2$, where $g$ is the SYM coupling constant.

Let us now specify the double scaling limit which one considers
on SYM side. We have \cite{Berenstein:2002jq} ${\cal N} =4$ SYM on
$S^3\times R$ with the action\footnote{The conformal
transformation from $R^4$ to $S^3\times R$ is:

\bqa
ds^2 = dx_\mu^2 = dr^2 + r^2 \, d\Omega_3, \quad |x_\mu| = e^t \Longrightarrow
ds^2 = e^{2t}\, \left(dt^2 + d \Omega_3\right). \nonumber
\eqa
Note that under such a conformal map the CFT correlation functions are converted into
transition amplitudes.}:

\bqa
S_{SYM} = \frac{1}{4\, g^2} \int d^4x \, {\rm Tr} \left\{F_{\mu\nu}^2 +
\left(D_\mu^{\phantom{\frac12}} \Phi_{\bar{I}}\right)^2
+ \left[\Phi_{\bar{I}}^{\phantom{\frac12}},
\, \Phi_{\bar{K}}\right]^2 + \frac23\, {\cal R} \, \Phi_{\bar{I}}^2
+ {\rm fermions} \right\}, \label{SYMac}
\eqa
where $\nu = 0,...,3$, ${\cal R}$ is the curvature of $S^3\times R$ (i.e. of $S^3$) and
$\Phi_{\bar{I}}$, $\bar{I} = 4,...,9$
are six real scalars. All the fields in the theory
are in the adjoint representation of the gauge group $U(N)$.
It is important that the ${\cal N} = 4$ SYM theory is conformally
invariant, because it has vanishing $\beta$-function. This explains the
conformal coupling of the scalars $\Phi$ to the curvature ${\cal R}$ in
\eq{SYMac}. Now it is clear what happens with the SYM theory in the limit (\ref{limit}).
One has to consider only those states of the theory on $S^3\times R$
which have $J\to \infty$ charge. To construct corresponding operators we combine two of the
six scalars $\Phi$ into:

\bqa
Z & = & \Phi_8 + {\rm i} \, \Phi_9. \label{comb}
\eqa
This $Z$ field carries unit of the $U_R(1)$ charge in question.
Hence, for example, the operator which saturates the
unitarity bound $\Delta = J$ is:

\bqa
{\cal O}^J \propto {\rm Tr} Z^J.
\eqa
The state corresponding to this operator plays the role of the vacuum in our
further considerations of the double scaling limit.

The proper excitations above such a vacuum should have finite
energies (anomalous conformal dimensions) in the double scaling limit (\ref{limit}).
According to \cite{Berenstein:2002jq} the simplest such states are as follows:

\bqa
{\cal O}^{J\, n}_{IK} \propto \sum_{l=0}^J {\rm Tr}\left\{\Phi_I \, Z^{l\phantom{\frac12}} \Phi_K \,
Z^{J-l}\right\} \, e^{\frac{2\, \pi\, {\rm i}\,l \, n}{J}}, \nonumber \\
{\rm where} \quad I = 4,...,7 \, .  \label{BMNop}
\eqa
In \cite{Constable:2002hw,Constable:2002vq,Kristjansen:2002bb,Beisert:2002bb} it was shown that
the one loop anomalous dimensions of these operators are
given by ($J\to\infty$):

\bqa
\Delta_n - J = 2 + \frac{g^2 \, N}{J^2} \, n^2 + \frac{1}{8 \pi^2} \frac{g^2 \,
J^2}{N}\left(\frac16 + \frac{35}{16\, \pi^2 \, n^2}\right) + \dots,
\eqa
where the second term on the RHS is the planar correction, while the third
one is non-planar. Actually it was argued in \cite{Gross:2002su,Santambrogio:2002sb} that all planar
corrections can be summed to give the expression:

\bqa
\Delta_n - J = 2 \sqrt{1 + \frac{g^2 \, N}{J^2} \, n^2} + \frac{1}{8 \pi^2} \frac{g^2 \,
J^2}{N}\left(\frac16 + \frac{35}{16\, \pi^2 \, n^2}\right) + \dots \, .
\label{states11}
\eqa
Under the identification (\ref{relation1}) one finds the relation between the spectra
of the SYM and ST $\Delta_n - J = m_{0, \,n}$, where $m_{0,\,n}$ is given by \eq{maphst}.
These formulae show that

\bqa
\lambda = \frac{g^2 \, N}{J^2} \label{relation2}
\eqa
plays the role of the effective t'Hooft coupling constant in the double
scaling limit. On the SYM side it is the weight for planar loop corrections to correlation functions
of the operators like (\ref{BMNop}). On the ST side, due to \eq{relation1}, it is the weight for
$\sigma$--model quantum corrections.

At the same time it is tempting to identify:

\bqa
g_s = \frac{J^2}{N}. \label{relation3}
\eqa
This means that the non-planar correction in \eq{states11} is
proportional to $\lambda \cdot g_s^2$.
Furthermore, below we show that

\bqa
{\cal R} \propto \mu^2, \label{relation4}
\eqa
where $\mu$ is the parameter of the pp--wave metric (\ref{metr2}).

  Thus, the free SYM is reproduced when

\bqa
\lambda \to 0, \quad \frac{J^2}{N} \to 0. \label{weak}
\eqa
On the ST side this limit corresponds to: $\mu \gg \frac{1}{\alpha' \, p^+}$
Hence, we can neglect the second term in \eq{LCact} in comparison with the
mass term for $x$:

\bqa
S_{free} = \frac{1}{2\pi} \, \int d\tau \, \int_0^{2\, \pi} d\sigma \left\{
\left(\pr_\tau x^{\phantom{\frac12}}_a(\sigma,\tau) \right)^{2}
- \mu^2 x_a^2(\sigma,\tau) + \bar{S}(\sigma,\tau)\, \left({\rm i}^{\phantom{\frac12}}
\pr_\tau + \mu \, \Gamma^{1234}\right)\, S(\sigma,\tau)\right\}, \label{SUSYosc}
\eqa
i.e. free SYM theory corresponds to the infinitely many (parameterized by $\sigma$)
non-interacting SUSY oscillator quantum mechanics\footnote{Note that in this
limit all the string excitations like (\ref{physst}) have
the same mass which is of the same order as the mass $\mu$ of oscillator
excitations in the pp--wave background.}. The relation between the free SYM and
the theory (\ref{SUSYosc}) can be established with the full mathematical
rigor. Moreover, it is not hard, using the
relations (\ref{relation1}), (\ref{relation2}), (\ref{relation3}) and (\ref{relation4}),
to define the SYM perturbation theory from the theory with the action (\ref{SUSYosc}).

The opposite --- strong coupling --- limit corresponds to the situation
when $\mu \ll \frac{1}{\alpha' \, p^+}$. Hence, $\pr_\sigma x_a = 0$, i.e. all the
string excitations become infinitely massive and decouple in the limit.
In this situation we obtain the weakly coupled supergravity
on the pp--wave background. It's scalar mode is described by the SUSY oscillator
quantum mechanics:

\bqa
S_{sc} = \frac{1}{2\pi} \, \int d\tau \, \left\{ \left(\pr_\tau
x_a^{\phantom{\frac12}}(\tau) \right)^{2}
- \mu^2 x_a^2(\tau) + \bar{S}(\tau)\, \left({\rm i}^{\phantom{\frac12}}
\hat{\pr}_\tau + \mu \, \Gamma^{1234}\right)\, S(\tau)\right\}. \label{SUSYosc1}
\eqa
It is easy to establish a map between physical states of the SYM theory in
the strong coupling limit (plus the double scaling limit) and physical
states of the theory (\ref{SUSYosc1}).
However, here we are not going to study this limit in any more detail.
From now on we consider SYM theory in the weak coupling limit (\ref{weak}) and tern
on first $\lambda$ and then $J^2/N$.

\subsection{From SYM to ST Hamiltonian}

In this subsection we carefully repeat all the arguments of
\cite{Berenstein:2002jq} and clearly specify the limit in which
they are valid. Consider separately all KK modes\footnote{This
does not mean that any of the KK modes are suppressed in any sense
in comparison with the zero modes. In fact, KK modes have masses
of the same order as the "zero--modes", i.e. of order of $\mu$.}
of the SYM fields on $S^3$. On the level of the zero-modes the
theory is described by MQM:

\bqa
S_{MQM} \propto \frac{N}{2}\, \int dt \, {\rm Tr} \left\{\left(D_0 A_i\right)^2 +
\left(D_0 \Phi_I\right)^2
+ \left|D_0 Z\right|^2 + \right. \nonumber \\ \left. +
g^2 \, N \, \left[A_i^{\phantom{\frac12}}, \, A_j\right]^2 +
g^2 \, N \, \left[\Phi_I^{\phantom{\frac12}}, \, A_i\right]^2 +
g^2 \, N \, \left|\left[Z^{\phantom{\frac12}}, \, \bar{Z}\right]\right|^2
+ \right. \nonumber \\ \left. +
g^2 \, N \, \left|\left[Z^{\phantom{\frac12}}, \, A_i\right]\right|^2 +
g^2 \, N \, \left|\left[Z^{\phantom{\frac12}}, \, \Phi_i\right]\right|^2
g^2 \, N \, \left[\Phi_I^{\phantom{\frac12}}, \, \Phi_K\right]^2 + \right. \nonumber \\
\left. + \frac23 \, {\cal R} \, \Phi_I^2 + \frac23 \, {\cal R} \, |Z|^2 + \frac23 \, {\cal R} \, A_i^2
+ {\rm fermions} \right\}, \label{MQM}
\eqa
where $D_0 = \pr_t + {\rm i} \, g^2 \, N \, A_0$ and
${\cal R}\, A_i^2$ term appears due to the SUSY with sixteen supercharges.

 It is worth mentioning at this point that everything which is done in this
subsection is valid only in the {\bf planar limit}. Thus, we have to
consider $J^2/N \to 0$. Then if we take $\lambda = g^2 \, N/ J^2 \to 0$ as well we
can neglect commutator terms
in \eq{MQM} if we are going to consider {\bf only} correlation functions of such
operators as (\ref{BMNop}) (which survive in the double scaling limit).
Hence, the MQM action in the extreme of the weak coupling limit is:

\bqa
\lim_{\lambda \to 0} S_{MQM} \propto N\, \int dt \, {\rm Tr} \left\{
\left(\pr_t A_i\right)^{2^{\phantom{\frac12}}} + \left(\pr_t \Phi_I\right)^2
+ \left|\pr_t Z\right|^2 + \right. \nonumber \\ \left. + \frac23 \, {\cal R} \, \Phi_I^2 +
\frac23 \, {\cal R} \, |Z|^{2^{\phantom{\frac12}}} + \frac23 \, {\cal R} \, A_i^2 +
{\rm fermions}\right\}. \label{freeMQM}
\eqa
This is true even despite the fact that $g^2 N \to \infty$ in our double
scaling limit.

It was shown in \cite{Gopakumar:1994iq} that in the {\bf planar approximation} the
master fields for $A_i$, $\Phi_I$ and $Z$ in the gaussian MQM are described
by (up to a gauge transformation \cite{Gopakumar:1994iq}):

\bqa
Z(t) & \Longrightarrow & \alpha^+ (t) \nonumber \\
\Phi_I (t) & \Longrightarrow &  \beta^+_{I}(t) , \nonumber \\
A_i(t) & \Longrightarrow & \beta^+_{i}(t),
\eqa
were $\alpha$ and $\beta$ are Cunz operators.
They obey no any relations except:

\bqa
\beta_{a}(t') \, \beta^+_{b} (t) = \delta_{ab} \, \delta(t' -
t), \nonumber \\
\alpha (t') \, \alpha^+ (t) = \delta(t' -
t), \nonumber \\
\beta_{a}(t') \, \alpha^+ (t) = 0,
\nonumber \\
\alpha (t') \, \beta^+_{a} (t) = 0, \nonumber \\
\alpha^+ \, \alpha + \sum_{a} \beta^+_{a} \beta_{a} = {\bf 1} -
|0\rangle\,\langle 0|, \label{rel}
\eqa
where $\beta_a = (\beta_I, \, \beta_i)$ and the vacuum $|0\rangle$ is
annihilated by all operators without $+$.

   If one considers in the MQM the states with a large $J$ charge
it is convenient to define new vacuum and operators:

\bqa
{\rm Tr} \, Z^J \, |0\rangle_{MQM} \Longrightarrow \left( \alpha^+\right)^J
\,|0\rangle \equiv |vac, \, J\rangle , \nonumber \\
\beta^+_{a\, l} \equiv \left(\alpha^+\right)^l \, \beta^+_a \,
\left(\alpha\right)^l, \quad l \leq J, \quad {\rm etc.}
\eqa
Similarly one can define $\alpha^+_l$.
The relations which are obeyed by these new operators are:

\bqa
\beta^+_{a\, l_1}(t) \, \beta^+_{b\, l_2}(t') = 0, \quad {\rm if} \quad l_2 < l_1,
\nonumber \\
\beta_{a\, l_1}(t) \, \beta_{b\, l_2}(t') = 0, \quad {\rm if} \quad l_2 > l_1,
\nonumber \\
\beta_{a\, l_1}(t) \, \beta^+_{b\, l_2}(t') = \delta_{l_1\, l_2}
\, \delta_{A\, B} \, \delta\left(t - t'\right), \eqa The question
is how to see Heisenberg algebra (algebra of the string oscillator
modes) within this bigger algebra. In \cite{Berenstein:2002jq} it
was argued that appropriate candidates for the free string
harmonics are the large $J$ limits of:

\bqa
b_n(t) = \frac{1}{\sqrt{J}} \, \sum^J_{l=0} \beta_l (t) \, e^{\frac{2\, \pi \, {\rm i} \, n \,
l}{J}} \Longrightarrow \frac{1}{\sqrt{J}} \, \sum^J_{l=0} \Phi_{[l]} (t) \,
e^{\frac{2\, \pi \, {\rm i} \, n \, l}{J}}. \label{relation}
\eqa
In fact, they obey the relations:

\bqa
\left[b_n(t), \, b_m(t')\right] = \frac{1}{J} \, \sum^J_{l\geq l'} \beta_l(t) \,
\beta_{l'}(t') \left(e^{\frac{2\, \pi \, {\rm i}\, (n\, l + m \, l')}{J}} -
e^{\frac{2\, \pi \, {\rm i}\, (n\, l' + m \, l)}{J}}\right), \nonumber \\
\left[b_n(t), \, b^+_m(t')\right] = \delta_{nm}\, \delta(t - t') -
\frac{1}{J} \, \sum^J_{l'=0} \, \sum^J_{l=0} \beta^+_{l'}(t) \,
\beta_l(t') \, e^{\frac{2\, \pi \, {\rm i}\, (n\, l - m \, l')}{J}}.
\eqa
In \cite{Berenstein:2002jq} it was argued that one should neglect the terms on the RHS
which have the pre--factor $1/J$. This is not completely clear for us. At this
point we do not know how to find any rigorous and explicit relation between
the two algebras in question. But we would like to point out that establishing a
relation between Heisenberg and Cunz algebras (which should exist) is very important for
the understanding of the SYM/pp--wave correspondence.

Assuming that the arguments of \cite{Berenstein:2002jq} are correct, it is easy to establish the
relation between the Hamiltonians of the MQM and ST on the pp--wave background.
In fact, let us consider deformations of the free MQM (\ref{freeMQM}) by the
potential terms of \eq{MQM}. In the large $J$ limit the contributions of
the most of the potential terms (to the correlation functions of the operators (\ref{BMNop}))
are suppressed. Only the terms Tr$|[Z, \, \Phi_I]|$ and Tr$|[Z, \, A_i]|$ do
contribute in the limit in question. Thus, substituting the master fields of
$Z$, $A$ and $\Phi$ into the Hamiltonian of the theory (\ref{MQM}) and using
the relations for the master fields, we obtain the Hamiltonian \cite{Berenstein:2002jq}:

\bqa
H_{BMN} \propto \int dt \, \left\{\sum_{a\, l}\left[\beta^+_{a\, l} \, \beta_{a\, l} +
g^2\, N \,
\left(\beta_{a\, l} + \beta^+_{a\, l} - \beta_{a\, l+1} - \beta^+_{a\, l+1}\right)^2
\right]\right\}. \label{BMNham}
\eqa
Now changing from $\beta$'s to $b$'s we obtain in the large $J$ limit:

\bqa
\lim_{J\to\infty} H_{BMN} \propto \frac12 \, \int_0^L ds \, \left\{\left(\pr_t \, x_a\right)^{2^{\phantom{\frac12}}} +
\left(\pr_s \, x_a\right)^2 + x_a^2 + {\rm fermions}\right\}, \quad L = J\, \sqrt{\frac{\pi}{g^2 \, N}},
\eqa
where $x$ is constructed from $b$'s as in \eq{expansion}.
Thus, in this way we obtain the LC Hamiltonian of the ST on the pp--wave
background.

  One can go even further \cite{Berenstein:2002jq} and establish
the correspondence between (gauge invariant) operators/states in MQM and
(physical) operators/states in the
ST. In fact, using the relation (\ref{relation}) and taking into account the
constraint (\ref{cons}) one can establish that:

\bqa\label{kind1}
\left|0^{\phantom{\frac12}}, p^+\right\rangle  &
\Longrightarrow & \lim_{J\to \infty} \, {\rm Tr} Z^J \,
\left|0^{\phantom{\frac12}}\right\rangle_{MQM} \Leftrightarrow \lim_{J\to \infty} \,
|vac^{\phantom{\frac12}}, J\rangle
\\ b^+_{n\,I}\, \tilde{b}^+_{n\, K} \,
\left|0^{\phantom{\frac12}},\, p^+\right\rangle
& \Longrightarrow & \lim_{J\to \infty} \, \sum_{l=0}^J {\rm Tr}\left\{\Phi_I \, Z^{l\phantom{\frac12}} \Phi_K \,
Z^{J-l}\right\} \, e^{\frac{2\, \pi\, {\rm i}\,l \, n}{J}} \,
\left|0^{\phantom{\frac12}}\right\rangle_{MQM}, \,\, {\rm etc.} . \nonumber
\eqa
Similar relations one can establish for the insertions of $A_i$ and $D_0$.
Curiously enough the MQM operators corresponding to the unphysical ST
states $b^+_{n}\, \tilde{b}^+_{m}\,
|0, \, p^+\rangle$ with $n \neq m$ do vanish \cite{Berenstein:2002jq}.

Moreover, only such SYM operators as in
(\ref{kind1}) have finite anomalous dimensions in the double
scaling limit (\ref{limit}). It was argued in \cite{Berenstein:2002jq} that all other kinds of SYM operators
(for example, with insertions od $\bar{Z}$) have divergent anomalous
dimensions in the double scaling limit (\ref{limit}), i.e. decouple
from OPE. In other words, corresponding
MQM states have divergent energies and decouple from the spectrum of the
theory in such a limit.

At this stage one could ask about KK modes of the SYM fields on $S^3$?
Are there any ST states related to them? To answer this question
consider the following operators in the SYM theory:

\bqa
{\cal O}_{\{m_1,m_2,\dots n_1,n_2,\dots\}} = {\rm Tr}\, \left\{Z^{l_1\phantom{\frac12}}
\Phi^{n_1}_{I_1} \, Z^{l_2} \,
\left(D_i^{m_1\phantom{\frac12}} Z^{l_3}\right) \dots \left(D_j^{m_2\phantom{\frac12}}
\Phi^{n_2}_{I_2}\right) \dots \right\}, \quad {\rm etc.}.
\eqa
The states corresponding to these operators are
linear combinations of the KK excitations of the SYM theory on $S^3$.
According to eq. (\ref{relation}) we can relate them to the
states in ST:

\bqa
b^+_{0\, A} \, |0, p^+ \rangle = |\dots 1, \, p^+ \rangle
& \Longrightarrow & \lim_{J\to \infty} \sum^J_{l=0} {\rm Tr}
\left\{Z^l \, (D_i Z)^{\phantom{\frac12}} Z^{J-l-1}\right\} \nonumber \\
b^+_{n\, i} \, \tilde{b}^+_{n, j}\, |0, p^+ \rangle
& \Longrightarrow & \lim_{J\to\infty} \sum^J_{l=0} {\rm Tr}
\left\{(D_i Z)^{\phantom{\frac12}} Z^{l-1} \, (D_j Z) \, Z^{J-l-1}\right\} \,
e^{\frac{2\, \pi\, {\rm i} \, l \, n}{J}}  \nonumber \\
b^+_{0\, i} \, b^+_{n\,I} \, \tilde{b}^+_{n\, K} \,|0, p^+ \rangle
& \Longrightarrow & \lim_{J\to\infty} \sum^J_{l_1, \, l_2} {\rm Tr}
\left\{\Phi_I \, Z^{l_1} \, \Phi_K \, Z^{l_2} \, (D_i Z)^{\phantom{\frac12}}
Z^{J- l_1 - l_2 - 1}\right\} \, e^{\frac{2\, \pi\, {\rm i} \, l_1 \, n}{J}},
{\rm etc.} \label{kind}
\eqa
where in the last line the differential $D_i$ in the sum acts both on $Z$
and $\Phi$'s. This is enforced by the commutation relations (\ref{commut}) for $b_0$ and
$b_n$, $n\neq 0$. Hence,
insertions of $D$'s into Tr$Z^J$ with extra phases are related to such
string excitations as $b^+_{n\, i} \, \tilde{b}^+_{n\, j} \, |0, \, p^+\rangle$.
Furthermore, insertions of $D$'s (as well as extra insertions of $\Phi$ without phase)
are related to the oscillator levels of the center of mass of the ST on
the pp-wave background. Below we give other arguments that support this identification:
in particular, we establish relations between correlation function of these "KK" operators
and scattering amplitudes of strings at excited oscillator levels.

In conclusion, there should be a correspondence between SYM theory
on $S^3\times R$ in the double scaling limit and the ST on the
pp--wave background. But from the presented considerations it is
not clear what kind of a relation is this? Is there any formula
like \eq{main} for this correspondence? Is this a relation between
planar SYM theory and free ST or is this a relation between
interacting theories? In the next sections we show that there is a
relation between the generating functional of correlation
functions of the operators (\ref{BMNop}) and generating functional
of LC transition amplitudes of the ST on the pp-wave background.

\subsection{Correlation functions}

Consider the tree--level value of the
two--point correlation function of the operators (\ref{BMNop}) in the
SYM theory on $S^3\times R$:

\bqa
\left\langle \overline{{\cal O}}^{J_1}_{n_1}\left(t_1,\, {\vec \phi}_1\right) \,
{\cal O}^{J_2}_{n_2}\left(t_2, \, {\vec \phi}_2\right)\right\rangle \propto
\delta_{J_1,\, J_2} \, \delta_{n_1,\, n_2} \,
\left(G\left(t_1,\, {\vec \phi}_1 \, |\, t_2, \, {\vec \phi}_2\right)\right)^{J_1+2} = \nonumber \\
= \delta_{J_1, \, J_2} \, \delta_{n_1,\, n_2} \,
\left(\sum_{L,\, k, \, m} C_L \, e^{{\rm i} \, \mu \, (L+1) \, (t_2 - t_1)} \,
Y_{L,\, k,\, m}({\vec \phi}_2 - {\vec \phi}_1)\right)^{J_1+2} = \nonumber \\
= \delta_{J_1,\, J_2} \, \delta_{n_1,\, n_2} \, \left[ C_0^{J_1+2}
\, e^{{\rm i} \, \mu \, (J_1+2) \, (t_2 - t_1)} + \sum_L
C_0^{J_1+1} \, C_L' \, e^{{\rm i} \, \mu \, (L + J_1 + 2) \, (t_2
- t_1)} \, \sum_{k, \, m} Y_{L,\, k,\, m}({\vec \phi}_2 - {\vec
\phi}_1) + \right. \nonumber \\ \left. + \sum_{L, \, L'} \,
C_0^{J_1} \, C_L' \, C_{L'}' \, e^{{\rm i} \, \mu \, (L + L' + J_1
+ 2) \, (t_2 - t_1)}\, \sum_{k,\, m,\, k', \, m'}  Y_{L,\, k,\,
m}({\vec \phi}_2 - {\vec \phi}_1) \, Y_{L',\, k',\, m'}({\vec
\phi}_2 - {\vec \phi}_1) + \dots\right], \label{twopo} \eqa where
we consider scalar part of the operator (\ref{BMNop}), ${\vec
\phi}$ is the coordinate on $S^3$, $\mu$ is the inverse
radius\footnote{We denote it as $\mu$ to show its relation to the
corresponding parameter of the pp--wave metric (\ref{metr2}).} of
$S^3$, $J_1 + 2$ is the classical conformal dimension of the
operator (\ref{BMNop}) and $G(t_1,\, {\vec \phi}_1 \, |\, t_2, \,
{\vec \phi}_2)$ is the scalar propagator\footnote{For definiteness
we use the retard propagator $(t_2 > t_1)$. All our considerations
below can be extended to other kinds of propagators on the both
sides of the relation.} of the theory \eq{SYMac} on $S^3\times R$.
Note that the first contribution in the third line is due to the
zero--modes on $S^3$.

In \eq{twopo} we have used the expansion of the propagator

\bqa
G(t,\, {\vec \phi}) = \sum_{L,\, k,\, m} \int \frac{dp}{2\, \pi}
\frac{e^{{\rm i} \, p \, t}}{\left(p^2 - \mu^2
\, L \, (L + 2) - \mu^2\right)} \, Y_{L,\, k,\, m} ({\vec \phi}) = \sum_{L,\, k, \, m} C_L \, e^{{\rm i} \,
\mu \, (L+1) \, t} \, Y_{L,\, k,\, m}({\vec \phi})
\eqa
in the spherical harmonics $Y$ on $S^3$. Here $L(L+2)$ is the eigen--value
corresponding to the $L$'s harmonic.

  Quantum corrections to \eq{twopo} within four--dimensional SYM theory were considered in
\cite{Berenstein:2002jq,Gross:2002su,Santambrogio:2002sb}. On the level of planar diagrams the result is given by the
shift of the conformal dimensions (\ref{states11}). Hence, in \eq{twopo} the
planar loop corrections are taken into account if we substitute

\bqa
J_1 + 2 \to J_1 + 2 \sqrt{1 + \frac{g^2 N}{J_1^2} \, n^2} \label{multilop}
\eqa
As well it is not hard to calculate quantum corrections to the first term in the
last line of the \eq{twopo} within MQM.
Amazingly enough the one loop result coincides with the one loop result
of the four--dimensional SYM theory. This is in accordance with the considerations of
\cite{Akhmedov:2002mi}. However, if the higher loop corrections within MQM to the zero mode
contribution in \eq{twopo} do not coincide with (\ref{multilop}) this is a
good sign in favor that it is four--dimensional SYM which corresponds to the ST on
the pp--wave background rather than MQM.

   To compare \eq{twopo} to string scattering amplitudes we have to properly normalize
the correlation function (\ref{twopo}). It is tempting to divide it
by the correlation function $\langle {\rm Tr} \overline{Z}^{J_1}(t_1) \, {\rm Tr} Z^{J_1}(t_2)\rangle$
to cancel the $J_1$ part in the value of the conformal dimension\footnote{It is interesting to know
whether the necessity of such a change of normalization will change conclusions of \cite{Beisert:2002bb} concerning
four--point functions.}.
After that, if we take into account quantum corrections and
make the Fourier transform over $t$ of each term in the last two lines of \eq{twopo} we obtain
that they are equal to:

\bqa
\delta_{J_1,\, J_2} \, \delta_{\Delta_1,\, \Delta_2} \,
\delta\left(p - 2\, \mu\, \sqrt{1 + \frac{g^2\, N}{J_1^2} \, n_1^2}\right) \nonumber \\
\delta_{J_1,\, J_2} \, \delta_{\Delta_1,\, \Delta_2} \,
\delta\left(p - \mu\, \left[L + 2 \, \sqrt{ 1 + \frac{g^2\, N}{J_1^2} \, n_1^2}\right]\right) \nonumber \\
\delta_{J_1,\, J_2} \, \delta_{\Delta_1,\, \Delta_2} \,
\delta\left(p - \mu\, \left[L + L' + 2 \, \sqrt{1 + \frac{g^2\, N}{J_1^2} \,
n_1^2}\right]\right), \quad {\rm etc.}.
\label{tpc}
\eqa
It is not hard to see that the first term (correlation function of the
zero KK harmonics on $S^3$) in this expression coincides with
the string scattering amplitude (\ref{stcorr1}), (\ref{stcorr2}) for the zero--oscillator harmonics on
pp--wave (if we identify $p=p^-$).

Furthermore, $SO(8)$ invariance acting on the $x_a$ and ${\vec M}$ is broken
to $SO(3)\times SO(5)$ by the fermions \cite{Metsaev:2001bj,Metsaev:2002re} in the theory. The $SO(3)$ factor
corresponds to the rotations of $S^3$ on SYM side. This is in accordance
with what we have discussed at the end of the subsection (3.1): the three among eight quantum numbers
$M_a$ are related to $L$, $k$ and $m$ of the spherical harmonics.
In fact, the last two expressions in \eq{tpc} are equal to the corresponding
correlation functions (\ref{stcorr1}), (\ref{stcorr2}), where only three $M_a$ are
turned on rather than all eight. Thus, we can reproduce full two--point
correlation function in SYM theory on $S^3 \times R$ from the ST on the pp--wave background\footnote{Note that
the operator mixing of \cite{Kristjansen:2002bb} is not important for us at this stage. First, we consider
planar approximation. Second, even if we were to take into account the mixing it would not
change our conclusions about space--time dependence of the correlation functions.}.

All our considerations so far establish relation between planar approximation of SYM
theory and free string theory. To make the first step towards establishing relation
between interacting theories we have to consider three--point functions of the zero
modes of the operators (\ref{BMNop}). For scalar parts of the operators (\ref{BMNop}) they are given by:

\bqa
\frac{\left\langle \overline{{\cal O}}^{J_1}_{n_1} (t_1) \, {\cal O}^{J_2}_{n_2} (t_2) \, {\cal O}^{J_3}_{n_3}
\right\rangle}{\left\langle {\rm Tr} \overline{Z}^{J_1} (t_1) \, {\rm Tr} Z^{J_2} (t_2)
\, {\rm Tr} Z^{J_3} (t_3)\right\rangle} \propto \nonumber \\ \propto \delta_{J_1, \, J_2 + J_3}
\, \delta_{\Delta_1, \Delta_2 + \Delta_3} \, C_{J_1\, J_2 \, J_3} \,
\exp\left\{{\rm i} \,\mu \, \frac{\tilde{\Delta}_1 + \tilde{\Delta}_2 -
\tilde{\Delta}_3}{2} \, (t_2 - t_1) \right\}\times \nonumber \\ \times
\exp\left\{{\rm i} \,\mu \, \frac{\tilde{\Delta}_1 + \tilde{\Delta}_3 -
\tilde{\Delta}_2}{2} \, (t_3 - t_1)\right\}\,
\exp\left\{{\rm i} \,\mu \, \frac{\tilde{\Delta}_2 + \tilde{\Delta}_3 -
\tilde{\Delta}_1}{2} \, (t_3 - t_2)\right\} = \nonumber \\ = \delta_{J_1, \, J_2 + J_3}
\, \delta_{\Delta_1, \Delta_2 + \Delta_3} \, C_{J_1\, J_2 \, J_3} \,
e^{- {\rm i} \,\mu \, \tilde{\Delta}_1 \, t_1}\, e^{{\rm i} \,\mu \, \tilde{\Delta}_2 \, t_2}\,
e^{{\rm i} \,\mu \, \tilde{\Delta}_3 \, t_3}\, \label{YMcorr}
\eqa
where $t_3 > t_2 > t_3$ and the full conformal dimensions of the operators (\ref{BMNop}) are given
by $\Delta = \tilde{\Delta} + J$. The structure constants $C_{J_1 \, J_2 \,
J_3}$ should be compared to the $\langle{\rm CFT}\rangle_3$ to find the full
agreement between this correlation function and (\ref{stcorr1}), (\ref{stcorr3}). At
this stage, however, we see that space--time dependencies agree. In fact,
after the Fourier transform of \eq{YMcorr} over $t_1$, $t_2$ and $t_3$ we
obtain that the three--point correlation function is equal to:

\bqa
\delta_{J_1, \, J_2 + J_3}
\, \delta_{\Delta_1, \Delta_2 + \Delta_3} \, C_{J_1\, J_2 \, J_3} \,
\delta\left(p_1 - \tilde{\Delta}_1\right) \,\delta\left(p_2 - \tilde{\Delta}_2\right) \,
\delta\left(p_3 - \tilde{\Delta}_3\right).
\eqa
This perfectly agrees with (\ref{stcorr1}), (\ref{stcorr3}) if $p=p^-$
and $\Delta$'s are given by \eq{multilop} (and if
the structure constants do coincide). Taking into account the agreement
between $\langle{\rm CFT}\rangle_3$ and structure constants $C_{J_1\, J_2\, J_3}$
found in \cite{Spradlin:2002rv} we can reproduce full four--dimensional
SYM correlation function on $S^3\times R$ from the ST. It would be
interesting, however, to reproduce the relation between
$\langle{\rm CFT}\rangle_3$ and $C_{J_1\, J_2\, J_3}$ within our approach.

\section{Conclusions and Acknowledgments}

In conclusion, we found the correspondence between the interacting
ST on the pp--wave background and the interacting SYM on $S^3\times R$ in the double scaling limit.
However, the explanation for this relation is still missing: our observations as well as those of
\cite{Berenstein:2002jq}, etc. still look mysterious for us.

  Apart from that there are several immediate questions answers to which are
not obvious for us. It is not obvious how it happens that superstring
theory torus correction does not vanish to reproduce non--planar correction
in \eq{states11}? It would be interesting to see whether the
operator mixing of \cite{Kristjansen:2002bb,Beisert:2002bb,Constable:2002hw,Constable:2002vq}
is somehow related to the fact that multiple--point SYM correlators are linear
combinations of the multiple--point ST correlators.

  Furthermore, one of the possible future directions is to calculate the following
correlation function:

\bqa
\frac{\left\langle vac , J_1 \right| \, \sum_{l_1} \beta_{l_1} \,
e^{\frac{2\, \pi \, {\rm i} \, l_1 \, n_1}{J_1}}\, \sum_{l_2} \beta_{l_2} \,
e^{- \frac{2\, \pi \, {\rm i} \, l_2 \, n_1}{J_1}} \, e^{{\rm i} \, H_{BMN} \, \Delta t} \,
\sum_{l_3} \beta^+_{l_3} \,
e^{\frac{2\, \pi \, {\rm i} \, l_3 \, n_2}{J_2}} \, \sum_{l_4} \, \beta^+_{l_4}
\, e^{\frac{- 2\, \pi \, {\rm i} \, l_4 \, n_2}{J_2}} \left|vac, J_2\right\rangle}{\left\langle vac , J_1 |
vac , J_1 \right\rangle}
\eqa
for finite $J$. This correlation function is the discretized analog of \eq{res}, where $H_{BMN}$
is given by \eq{BMNham}. Hence, its should give a regularization of the string theory amplitudes.

I would like to acknowledge discussions with A.Mikhailov,
A.Tseytlin, A.Gorsky, A.Marshakov, G.Semenoff, M.Staudacher and especially with
N.Ishibashi and A.Gerasimov. I would like to thank Kitazawa san,
Iso san, Kawai san, Yoneya san, Hashimoto san, Osayuki san, Sugawara san and especially Ishibashi san,
Tada san and Suzuki san for hospitality in
KEK, RIKEN, Kanazawa Univ. and Hongo and Komaba Tokyo Univ. where this project was initiated.
This work was partially supported by the grant RFBR 02-02-17260 and
INTAS-00-390.


\begin{thebibliography}{50}

\bibitem{Morozov:hh}
A.~Morozov,
Phys.\ Usp.\  {\bf 37}, 1 (1994)
[arXiv:hep-th/9303139].

\bibitem{Witten:1988hc}
E.~Witten,
Nucl.\ Phys.\ B {\bf 311}, 46 (1988).

\bibitem{Banks:1996vh}
T.~Banks, W.~Fischler, S.~H.~Shenker and L.~Susskind,
Phys.\ Rev.\ D {\bf 55}, 5112 (1997)
[arXiv:hep-th/9610043].

\bibitem{Ishibashi:1996xs}
N.~Ishibashi, H.~Kawai, Y.~Kitazawa and A.~Tsuchiya,
Nucl.\ Phys.\ B {\bf 498}, 467 (1997)
[arXiv:hep-th/9612115].

\bibitem{Maldacena:1997re}
J.~M.~Maldacena,
Adv.\ Theor.\ Math.\ Phys.\  {\bf 2}, 231 (1998)
[Int.\ J.\ Theor.\ Phys.\  {\bf 38}, 1113 (1999)]
[arXiv:hep-th/9711200].

\bibitem{Gubser:1998bc}
S.~S.~Gubser, I.~R.~Klebanov and A.~M.~Polyakov,
Phys.\ Lett.\ B {\bf 428}, 105 (1998)
[arXiv:hep-th/9802109].

\bibitem{Witten:1998qj}
E.~Witten,
Adv.\ Theor.\ Math.\ Phys.\  {\bf 2}, 253 (1998)
[arXiv:hep-th/9802150].

\bibitem{Aharony:1999ti}
O.~Aharony, S.~S.~Gubser, J.~M.~Maldacena, H.~Ooguri and Y.~Oz,
Phys.\ Rept.\  {\bf 323}, 183 (2000)
[arXiv:hep-th/9905111].

\bibitem{Akhmedov:un}
E.~T.~Akhmedov,
Phys.\ Usp.\  {\bf 44}, 955 (2001)
[Usp.\ Fiz.\ Nauk {\bf 44}, 1005 (2001)].

\bibitem{Akhmedov:1999rc}
E.~T.~Akhmedov,
arXiv:hep-th/9911095.

\bibitem{Berenstein:2002jq}
D.~Berenstein, J.~M.~Maldacena and H.~Nastase,
JHEP {\bf 0204}, 013 (2002)
[arXiv:hep-th/0202021].

\bibitem{Akhmedov:2002gq}
E.~T.~Akhmedov,
arXiv:hep-th/0202055.

\bibitem{Metsaev:2001bj}
R.~R.~Metsaev,
Nucl.\ Phys.\ B {\bf 625}, 70 (2002)
[arXiv:hep-th/0112044].

\bibitem{Metsaev:2002re}
R.~R.~Metsaev and A.~A.~Tseytlin,
Phys.\ Rev.\ D {\bf 65}, 126004 (2002)
[arXiv:hep-th/0202109]. \\

\bibitem{Tseytlin} A.Tseytlin, private communications.

\bibitem{Gross:2002su}
D.~J.~Gross, A.~Mikhailov and R.~Roiban,
Annals Phys.\  {\bf 301}, 31 (2002)
[arXiv:hep-th/0205066].

\bibitem{Santambrogio:2002sb}
A.~Santambrogio and D.~Zanon,
Phys.\ Lett.\ B {\bf 545}, 425 (2002)
[arXiv:hep-th/0206079].

\bibitem{Constable:2002hw}
N.~R.~Constable, D.~Z.~Freedman, M.~Headrick, S.~Minwalla, L.~Motl, A.~Postnikov and W.~Skiba,
JHEP {\bf 0207}, 017 (2002)
[arXiv:hep-th/0205089].

\bibitem{Spradlin:2002rv}
M.~Spradlin and A.~Volovich,
arXiv:hep-th/0206073.

\bibitem{Akhmedov:1998vf}
E.~T.~Akhmedov,
Phys.\ Lett.\ B {\bf 442}, 152 (1998)
[arXiv:hep-th/9806217].

\bibitem{Berenstein:2002sa}
D.~Berenstein and H.~Nastase,
arXiv:hep-th/0205048.

\bibitem{Blau:2002dy}
M.~Blau, J.~Figueroa-O'Farrill, C.~Hull and G.~Papadopoulos,
Class.\ Quant.\ Grav.\  {\bf 19}, L87 (2002)
[arXiv:hep-th/0201081].

\bibitem{Gopakumar:1994iq}
R.~Gopakumar and D.~J.~Gross,
Nucl.\ Phys.\ B {\bf 451}, 379 (1995)
[arXiv:hep-th/9411021].

\bibitem{Kristjansen:2002bb}
C.~Kristjansen, J.~Plefka, G.~W.~Semenoff and M.~Staudacher,
Nucl.\ Phys.\ B {\bf 643}, 3 (2002)
[arXiv:hep-th/0205033].

\bibitem{Beisert:2002bb}
N.~Beisert, C.~Kristjansen, J.~Plefka, G.~W.~Semenoff and M.~Staudacher,
arXiv:hep-th/0208178.

\bibitem{Constable:2002vq}
N.~R.~Constable, D.~Z.~Freedman, M.~Headrick and S.~Minwalla,
JHEP {\bf 0210}, 068 (2002)
[arXiv:hep-th/0209002].


\bibitem{Akhmedov:2002mi}
E.~T.~Akhmedov and A.~V.~Smilga,
arXiv:hep-th/0202027.

\end{thebibliography}
\end{document}